\documentstyle[epsfig,prl,aps,amsfonts,twocolumn]{revtex}

\begin{document}
\title{\hfill \\
A Hartree-Fock Study of Persistent Currents in Disordered Rings}
\author{A. Kambili\thanks{e-mail: a.kambili@lancaster.ac.uk}, C. J. Lambert, } 
\address{School of Physics and Chemistry, Lancaster University Lancaster LA1 4YB, UK}
\author{J. H. Jefferson} 
\address{Defence Evaluation and Research Agency, Electronic and Optical Material
Centre, Malvern, Worcs. WR14 3PS UK}

\date{\today}
\maketitle
 
\begin{abstract}
For a system of spinless fermions in a disordered mesoscopic ring, interactions 
can give rise to an enhancement of the persistent current by orders of magnitude.
The increase in the current is associated with a charge reorganization of the 
ground state. The interaction strength for which this reorganization takes place
is sample-dependent and the log-averages over the ensemble are not representative.
In this paper we demonstrate that the Hartree-Fock method closely reproduces
results obtained by exact diagonalization. For spinless fermions subject to a
short-range Coulomb repulsion U we show that due to charge reorganization the
derivative of the persistent current is a discontinuous function of U. Having 
established that the Hartree-Fock method works well in one dimension, we present
corresponding results for persistent currents in two coupled chains. \\
Pacs numbers: 71.30.+h, 05.45.+b, 71.10.Pm, 72.15.Rn  \\
\end{abstract}


\par In a normal-metal mesoscopic ring threaded by a magnetic flux \cite{1}, measured values of the 
persistent current \cite{2} are two orders of magnitude larger than predicted. This result
suggests that a quantitative theory must treat electron-electron interactions and disorder on
an equal footing. Previous studies of spinless fermions concentrated on the behaviour of ensemble
averages of the persistent current and led to the conclusion that repulsive
interactions cannot significally enhance the current. It was therefore argued that only
systems which include spin could show such an increase of the current \cite{3}. However,
more recently it was shown \cite{4} that for one-dimensional systems of spinless fermions 
interacting through a short-range Coulomb repulsion U, the persistent current does increase. This
enhancement of the current is accompanied by a charge reorganization of the ground
state which happens at different values of the interaction strength U, depending
on the disorder realization, and therefore the ensemble averaged persistent current may not be
relevant.
\par The results of \cite{4} were obtained using the density matrix renormalization group
(DMRG) technique \cite{5} which is essentially exact and contains correlation effects. However, this 
technique is computationally demanding and not easily extended to higher dimensions. Therefore it is 
of interest to determine whether or not the results of \cite{4} are
contained in a mean-field description using a single Slater determinant. In this paper we address
this question, by presenting results obtained using the Hartree-Fock method for spinless fermions
interacting via a short-range Coulomb repulsion. In one dimension we show 
that the Hartree-Fock method agrees with the exact results of \cite{4}, reproducing
the  predicted behaviour of the persistent current as well as the sample-dependent charge 
reorganization. After establishing the validity of the method, we extend the calculation
to a quasi-one dimensional system comprising two parallel chains.
\par The total Hamiltonian for a system of $N$ spinless fermions on a
disordered chain ($1$D) of $M$ sites is

\begin{equation}
H=\sum _{i=1}^{M}\varepsilon_{i}c_{i}^{\dag}c_{i}-\sum _{i,j=1}^{M}
t_{ij}c_{i}^{\dag}c_{j}+\frac{1}{2}\sum _{i,j=1}^{M}U_{ij}c_{i}^{\dag}c_{i}
c_{j}^{\dag}c_{j}
\end{equation}

The operators $c_{i}^{\dag}$ and $c_{i}$ are creation and annihilation operators
for an electron on site $i$, the on-site energies $\varepsilon_{i}$ are random
variables uniformly distributed over the interval $-\frac{W}{2}$ to $+\frac{W}{2}$ and 
$U_{ij}$ is a nearest-neighbour interaction of the form

\begin{equation}
U_{ij}=\left\{ \begin{array}{ll}
                U & \mbox{if $j=i\pm 1$} \\
	        0 & \mbox{otherwise}
\end{array}\right.
\end{equation}

The hopping elements $t_{ij}$ are restricted to nearest neighbours with $t_{i,i\pm1}=t=1$
except at the ends of the chain, for which $t_{1N}=t_{N1}=1$ (periodic boundary conditions) or 
$t_{1N}=t_{N1}=-1$  (anti-periodic boundary conditions).

\begin{figure}[h]
\centerline{\epsfig{figure=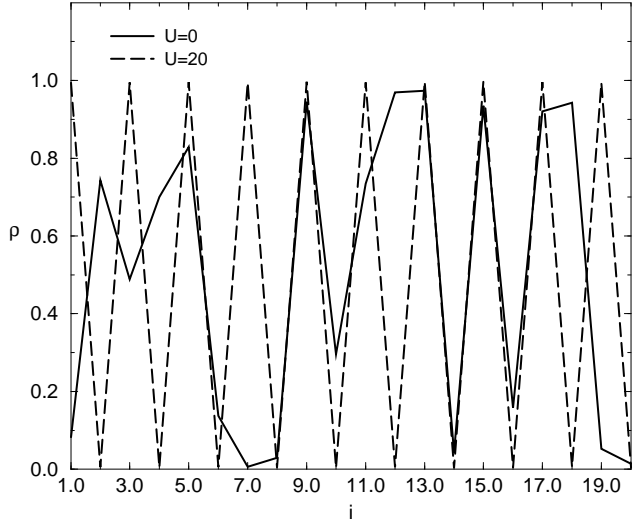,width=10cm}}
\caption{Charge density for a sample with $N=10$ particles on a chain (1D) with $M=20$ sites
for free fermions and for $U=0$ and $U=20$ }
\label{1}
\end{figure}

The single-particle Hartree-Fock equation corresponding to equation (\ref{1}) is of the
form

\begin{eqnarray}
\varepsilon_{i}\Psi^{n}(i)-t_{ii-1}\Psi^{n}(i-1)-t_{ii+1}\Psi^{n}(i+1)\nonumber \\
+\sum_{m=1}^{N}\sum_{j=1}^{M}|\Psi^{m}(j)|^{2}U_{ij}\Psi^{n}(i)- \nonumber \\
\sum_{m=1}^{N}\sum_{j=1}^{M}\Psi^{\ast m}(j)\Psi^{m}(i)U_{ij}\Psi^{n}(j)=
E_{n}\Psi^{n}(i)
\end{eqnarray}

where $\Psi^{n}(i)$ is the amplitude of the $n$th single-particle wavefunction on site $i$.
The third and fourth terms are the direct (Hartree) and
exhange (Fock) potentials, respectively. 

\par The parameters $N$, $M$, $t$, $W$, and $U$ were chosen to be exactly the same as in
\cite{4} so as to directly test the accuracy of the method. Here we present results
for $M=20$ sites and $N=10$ particles (half-filling) and the strength of the disorder
was taken to be large, $W=9$.
 At zero $U$ the charge density of a given sample is spatially inhomogeneous due to
the presence of such disorder. As $U$ increases we observe a reorganization of 
the charge, and for large $U$, obtain a homogeneous configuration in which
the particles are equally spaced. This reorganization happens at different
values of the interaction strength for different samples. As an example, figure \ref{1}
shows the charge density for a single sample in the free electron case ($U=0$) and for $U=20$,
which is large enough to yield a periodic array of charges.

\par In addition to the charge density we have studied the phase sensitivity $D$,
which is a measure of the delocalization effect mentioned above and is defined by

\begin{equation}
D(U)=\frac{M}{2}\Delta E
\end{equation}

Here $\Delta E=(-1)^{N}(E_{g}(0)-E_{g}(\pi))$ is the difference in the ground state energy between
periodic and anti-periodic boundary conditions, where the Hartree-Fock ground state energy is

\begin{equation}
E_{g}=\frac{1}{2}\sum_{n=1}^{N}[E_{n}+\sum_{i,j=1}^{M}\Psi^{\ast n}(i)h_{ij}\Psi^{n}(j)]
\end{equation}

with $h_{ij}=\varepsilon_{i}\delta _{ij}+t_{ij}\delta_{ij\pm 1}$.

\par
\begin{figure}[h]
\centerline{\epsfig{figure=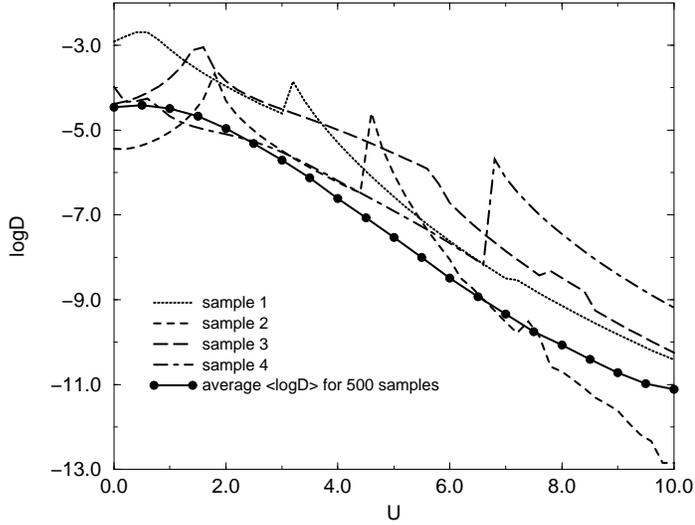,width=10cm}}
\caption{Phase sensitivity for half-filling ($M=20$, $N=10$) in one dimension for four samples.
The average has been obtained for 500 disorder realizations}
\label{2}
\end{figure}

\begin{figure}[h]
\centerline{\epsfig{figure=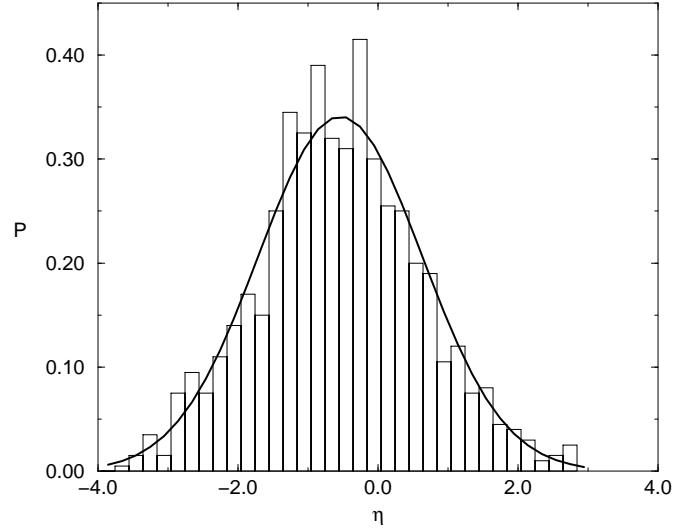,width=10cm}}
\caption{Probability distribution of $\eta=\log{D}(2)-\log{D}(0)$ in 1D calculated from
1000 samples. The curve is fitted by a Gaussian distribution with $<\eta>=-0.53$ and
$\sigma^{2}=1.37$}
\label{3}
\end{figure}

 In agreement with \cite{4} we find peaks in $\log{D}$ at sample-dependent
values of $U$, associated with reorganization of the ground-state
charge density. For positive $U$, figure \ref{2} shows the ensemble average of $\log{D}$
along with results for four individual samples. For negative $U$ the mean field
equations do not converge and no results were obtainable. Fig. \ref{2} is in remarkable
agreement with the exact results of \cite{4}. For example the average of $\log{D}$, exhibits 
a local maximum around $U\approx t$. Following \cite{4} we have also examined the relative increase of the
phase sensitivity with respect to the free fermion case, $\eta=\log{D}(U)-\log{D}(0)$, and in
agreement with \cite{4}, obtain a log-normal distribution for $\eta$ (figure \ref{3}).

\par
\begin{figure}[h]
\centerline{\epsfig{figure=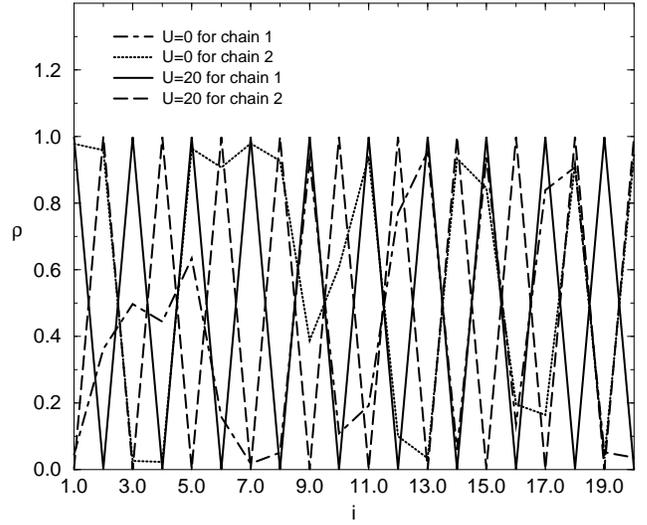,width=10cm}}
\caption{Charge density for a sample with $N=20$ particles in two rings (2D) with $M=20$ sites
in each ring for free fermions and for $U=0$ and $U=20$}
\label{4}
\end{figure}

 Having established the validity of Hartree-Fock theory as a method for computing the 
charge density and phase sensitivity of one-dimensional rings, we now extend our analysis to 
two coupled one-dimensional chains. 
For a system consisting of two rings with $20$ sites in each, and $20$ spinless 
fermions in total, we again examine the case of strong disorder, $W=9$. In
figure \ref{4} we show the charge density for the two rings. As we can see, the charge
reorganization that was present in $1$D is also obvious for two chains. For very strong
$U$ the particles localize in the odd and the even sites of the first and second chains
respectively, as expected classically. Thus, one again obtains 
the delocalization effect associated with the crossover from an Anderson to a Mott insulator. 
\par
\begin{figure}[h]
\centerline{\epsfig{figure=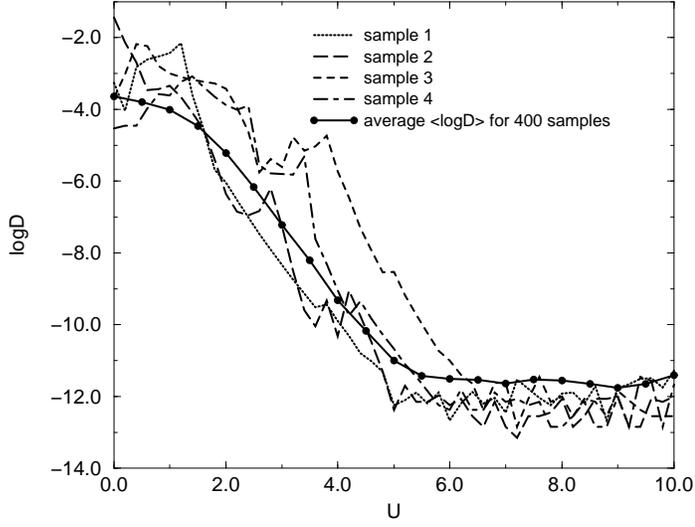,width=10cm}}
\caption{Phase sensitivity for half-filling ($M=40$, $N=20$) in two dimensions for four samples.
The average has been obtained from 400 disorder realizations}
\label{5}
\end{figure}

\begin{figure}[h]
\centerline{\epsfig{figure=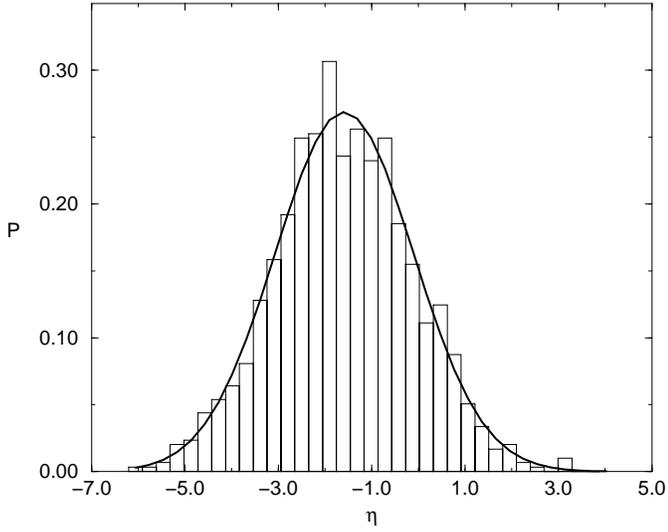,width=10cm}}
\caption{Probability distribution of $\eta=\log{D}(2)-\log{D}(0)$ in 2D calculated from
1000 samples. $<\eta>=-1.59$ and $\sigma^{2}=2.2$}
\label{6}
\end{figure}

 The phase sensitivity and the probability distribution of the relative increase $\eta$ are
shown in figures \ref{5} and \ref{6}, which again demonstrate that interactions produce an increase in 
the fluctuations of the current. However, in contrast with a single chain, the average of $\log D$ no
longer possesses a local maximum and instead decreases monotonically with increasing $U$.

\par In summary, the above results demonstrate that the effects discussed in \cite{4} are
contained in a single Slater-determinant ground state and are describable by mean field
Hartree-Fock theory. By extending the analysis to two chains, we find that the 
maximum in the average of $\log D$ is no longer present, which suggests that this feature
may be a peculiarity of strictly one-dimensional systems. The fact that Hartree-Fock theory is
applicable in one dimension, where mean field theories are least accurate, indicates that in
higher dimensions Hartree-Fock theory should be sufficient to describe the ground state of a system
with electron-electron interactions and disorder, at least in the strong disorder limit.


\end{document}